# Mutual Information Games in Multi-user Channels with Correlated Jamming[*]

Shabnam Shafiee       Sennur Ulukus

Department of Electrical and Computer Engineering

University of Maryland, College Park, MD 20742

*sshafiee@umd.edu*       *ulukus@umd.edu*

August 6, 2018

## Abstract

We investigate the behavior of two users and one jammer in an AWGN channel with and without fading when they participate in a non-cooperative zero-sum game, with the channel's input/output mutual information as the objective function. We assume that the jammer can eavesdrop the channel and can use the information obtained to perform correlated jamming. We also differentiate between the availability of perfect and noisy information about the user signals at the jammer. Under various assumptions on the channel characteristics, and the extent of information available at the users and the jammer, we show the existence, or otherwise non-existence of a simultaneously optimal set of strategies for the users and the jammer. In all the cases where the channel is non-fading, we show that the game has a solution, and the optimal strategies are Gaussian signalling for the users and linear jamming for the jammer. In fading channels, we envision each player's strategy as a power allocation function over the channel states, together with the signalling strategies at each channel state. We define sub-games at each channel state, given any set of power allocation functions for the players. Based on our results in the non-fading channels, the sub-games always have a solution which is Gaussian signalling for the users and linear jamming for the jammer. Given the solution of the sub-games, we reduce the game solution to a set of power allocation functions for the players and show that when the jammer is uncorrelated, the game has a solution, but when the jammer is correlated, a set of simultaneously optimal power allocation functions for the users and the jammer does not always exist. In this case, we characterize the max-min user power allocation strategies and the corresponding jammer power allocation strategy.

---

[*]This work was supported by NSF Grants ANI 02-05330, CCR 03-11311, CCF 04-47613 and CCF 05-14846; and ARL/CTA Grant DAAD 19-01-2-0011, and presented in part at the Conference on Information Sciences and Systems, Baltimore, MD, March 2005 [1] and the Military Communication Conference, Atlantic City, NJ, October 2005 [2].

# 1 Introduction

Correlated jamming, the situation where the jammer has full or partial knowledge about the user signals has been studied in the information-theoretic context under various assumptions [3–5]. In [3] the best transmitter/jammer strategies are found for an AWGN channel with one user and one jammer who participate in a two person zero-sum game with the mutual information as the objective function. The jammer is power constrained and has full or partial knowledge of the transmitted signal which may be obtained through eavesdropping. In [4], the problem is extended to a single-user MIMO fading channel with the assumption that the jammer has full knowledge of the user signal. This model has been further extended in [5] to consider fading in the channel between the jammer and the receiver. In [5] various assumptions are made on the availability of the user channel state at the user, and the jammer channel state at the jammer.

In this paper, we study a *multi-user* system under correlated jamming. Without loss of generality, we consider a system of two users and one jammer who has full or partial knowledge of the user signals through eavesdropping, and examine the existence of optimum user and jammer strategies towards achieving maximum mutual information. In the non-fading two user channel, we show that the game has a solution which is Gaussian signalling for the users, and linear jamming for the jammer. Here we define linear jamming as employing a linear combination of the available information about the user signals plus Gaussian noise, where the available information is the user signals in the case of perfect information, and a noisy version of the user signals in the case of eavesdropping. We show that the power that the jammer allocates for jamming each user's signal is proportional to that user's power.

We then consider fading in the user channels. As opposed to [5], where the user channel states could only be known at the users, we assume the possibility of the jammer gaining information about the user channel states by eavesdropping the feedback channel from the receiver to the users. We show that if the jammer is not aware of the user channel states, it would disregard its eavesdropping information and only transmit Gaussian noise. If the jammer knows the user channel states but not the user signals, the game has a solution which is composed of the optimal user and jammer power allocation strategies over the channel states, together with Gaussian signalling and linear jamming at each channel state. The optimal power allocations in this case are such that only one user transmits at any given channel state. If the jammer knows the user channel states and the user signals, the game does not always have a Nash equilibrium solution, in which case, we characterize the max-min user strategies, and the corresponding jammer best response. The max-min user strategy corresponds to the user's best move, in a situation where the user chooses its strategy only once, while after the user chooses its strategy, the jammer can observe it and choose the corresponding best jamming. Note that if the game had a solution, max-min and min-max strategies would have been the same, and would also be the same as the game solution.

The term capacity will hereafter always refer to the channel's information capacity, which is defined as the channel's maximum input/output mutual information [7].

## 2 System Model

Figure 1 shows a communication system with two users and one jammer. We consider several different settings based on the channel characteristics and the jammer's information. In the absence of fading, the attenuations of the user channels are known to everyone. Therefore, we can assume that the attenuations are scalars. The AWGN channel with two users and one jammer is modelled as

$$Y = \sqrt{h_1}X_1 + \sqrt{h_2}X_2 + \sqrt{\gamma}J + N \tag{1}$$

where $X_i$ is the $i^{th}$ user's signal, $h_i$ is the attenuation of the $i^{th}$ user's channel, $J$ is the jammer's signal, $\gamma$ is the attenuation of the jammer's channel and $N$ is a zero-mean Gaussian random variable with variance $\sigma_N^2$. To model fading in the received powers, we consider $h_i$ and $\gamma$ as fading random variables, and to further model the phase of the channel coefficients, we substitute the scalar attenuations $\sqrt{h_i}$ and $\sqrt{\gamma}$, with complex fading random variables $H_i$ for the amplitude fading coefficient of the $i^{th}$ user's channel, and $\Gamma$ for the amplitude fading coefficient of the jammer's channel

$$Y = H_1 X_1 + H_2 X_2 + \Gamma J + N \tag{2}$$

All fading random variables are assumed to be i.i.d. The user and jammer power constraints are

$$E[X_i^2] \leq P_i, \quad i = 1, 2 \tag{3}$$
$$E[J^2] \leq P_J \tag{4}$$

Regarding the knowledge of the jammer about the transmitted signals, we analyze both cases of perfect information and imperfect information gained through eavesdropping. In the first case, we assume that the jammer knows the signals of the users perfectly, i.e., it knows $X_1$ and $X_2$ at the beginning of its transmission. In the second case, we assume an AWGN eavesdropping channel for the jammer

$$Y_e = \sqrt{g_1}X_1 + \sqrt{g_2}X_2 + N_e \tag{5}$$

where $Y_e$ is the signal received at the jammer, $g_i$ is the attenuation of the $i^{th}$ user's eavesdropping channel and $N_e$ is a zero-mean Gaussian random variable with variance $\sigma_e^2$. Therefore, in this case, the jammer knows a noisy version of a linear combination of the user signals, $X_1$ and $X_2$. To model fading in the received powers, we consider $\sqrt{g_i}$ as real fading variables, and to model fading in the received amplitudes, we substitute them with complex amplitude fading random variables $G_i$. The receiver is assumed to know the user channel states, while various assumptions are made on the amount of information that the users and the jammer have about the channel fading realization of the communication and eavesdropping channels; these assumptions are stated at the beginning of each subsection.

## 3 Jamming in Non-fading Multi-user AWGN Channels

In this section, we find the best user/jammer strategies when the channels are non-fading, both when the jammer knows the exact user signals, and when it eavesdrops the users' channel and obtains a noisy version of a linear combination of the user signals.

### 3.1 Jamming with Complete Information

Here the system model is (1) where the attenuations are constant scalars, and $X_1$ and $X_2$ are known to the jammer. The jammer and the two users are involved in a zero-sum game with the input/output mutual information as the objective function. We investigate the existence and uniqueness of a Nash equilibrium solution for this game [8]. A Nash equilibrium is a combination of strategies, one for each player, such that no player has an incentive for unilaterally changing its own strategy, meaning that, no player will gain more, by unilaterally deviating from the Nash equilibrium solution. Note that in a zero-sum game, if the objective function is convex over the set of the strategies of the players that are minimizing it, and concave over the set of the strategies of the players that are maximizing it, then the mathematical saddle point of the objective function corresponds to the game's Nash equilibrium solution. However, in general, all mathematical saddle points of an objective function may not necessarily correspond to a game solution, e.g., matrix games [8].

The arguments in [7, Theorem 2.7.4] can be easily extended to the two user system to show that if $(X_1, X_2, Y) \sim f(x_1)f(x_2)f(y|x_1, x_2)$, the input/output mutual information $I(X_1, X_2; Y)$ is a concave function of $f(x_1)$ for fixed $f(x_2)$ and $f(y|x_1, x_2)$, a concave function of $f(x_2)$ for fixed $f(x_1)$ and $f(y|x_1, x_2)$, and a convex function of $f(y|x_1, x_2)$ for fixed $f(x_1)$ and $f(x_2)$. Due to the convexity/concavity of the mutual information with respect to the channel transition probability distribution/user input probability distribution, and given that the set of the user and jammer signallings which satisfy the corresponding power constraints is convex, $I(X_1, X_2; Y)$ has a saddle point in that set, which is the Nash equilibrium solution of the game [9, Theorem 16, page 75], [10, Proposition 2.6.9]. In the sequel, we show that when the users employ Gaussian signalling, the best jamming strategy is linear jamming (linear combination of the user signals plus Gaussian noise), and when the jammer employs linear jamming, the best strategy for the users is Gaussian signalling, which proves that Gaussian input distributions for the users and linear jamming for the jammer is a saddle point of the input/output mutual information, and therefore, a Nash equilibrium solution for the game. Due to the interchangeability property of game solutions [9, Theorem 8, page 48], if there is any other pair of strategies which is a game solution as well, it has to result in the same mutual information value as the game solution corresponding to Gaussian signaling and linear jamming [9, Theorem 7, page 48].

First, assume that the jammer employs linear jamming

$$J = \rho_1 X_1 + \rho_2 X_2 + N_J \tag{6}$$

The power constraint on the jammer will force the following condition

$$\rho_1^2 P_1 + \rho_2^2 P_2 + \sigma_{N_J}^2 \leq P_J \tag{7}$$

Using (1), the output of the channel will be

$$Y = (\sqrt{h_1} + \sqrt{\gamma}\rho_1)X_1 + (\sqrt{h_2} + \sqrt{\gamma}\rho_2)X_2 + \sqrt{\gamma}N_J + N \tag{8}$$

From the users' perspective, the channel becomes an AWGN multiple access channel, and therefore the best signalling scheme for the users is Gaussian [7].

Next, we should show that if the users perform Gaussian signalling, then the best jamming strategy is linear jamming. The channel output is as in (1), where $X_1$ and $X_2$ are independent Gaussian random variables, and $J$, the jammer signal, is an arbitrary random variable to be chosen by the jammer. We write the input/output mutual information of the channel

$$I(Y; X_1, X_2) = h(X_1, X_2) - h(X_1, X_2|Y) \tag{9}$$

The jammer's strategy can only affect the second term above. We develop a sequence of upper bounds on the second term,

$$h(X_1, X_2|Y) = h(X_1 - a_1 Y, X_2 - a_2 Y|Y) \tag{10}$$
$$\leq h(X_1 - a_1 Y, X_2 - a_2 Y) \tag{11}$$
$$\leq \frac{1}{2} \log\left((2\pi e)^2 |\mathbf{\Lambda}|\right) \tag{12}$$

where $\mathbf{\Lambda}$ is the covariance matrix of $(X_1 - a_1 Y, X_2 - a_2 Y)$. The inequalities hold for arbitrary $a_1$ and $a_2$. We choose $a_1 = E[X_1 Y]/E[Y^2]$ and $a_2 = E[X_2 Y]/E[Y^2]$. We now prove the optimality of linear jamming in two steps. We first consider the set of all jamming signals which result in the same $\boldsymbol{\Lambda}$, and show that if this set includes a linear jammer, then that linear jammer is the optimal jammer over this set. Then, we consider the set of all feasible jamming signals and show that for any jamming signal in this set, there exists a linear jammer in this set resulting in the same $\boldsymbol{\Lambda}$. Here, the feasibility is in the sense of the jammer's available power.

Consider the set of all jamming signals which result in the same $\boldsymbol{\Lambda}$ in (12). Assume that there is a linear jamming signal in this set. This jamming signal is jointly Gaussian with $X_1$ and $X_2$, hence $X_1 - a_1 Y$, $X_2 - a_2 Y$ and $Y$ are jointly Gaussian. Moreover, $a_1$ and $a_2$ are chosen such that $X_1 - a_1 Y$ and $X_2 - a_2 Y$ are uncorrelated with $Y$, therefore, since they are all Gaussian, $X_1 - a_1 Y$ and $X_2 - a_2 Y$ are independent of $Y$. We conclude that this jamming signal achieves both (11) and (12) with equality, therefore, it is optimal over this set.

Now we show that any $\boldsymbol{\Lambda}$ achievable by any feasible jamming signal, is also achievable by a feasible linear jamming signal. For the chosen values of $a_1$ and $a_2$, $\boldsymbol{\Lambda}$ is

$$\boldsymbol{\Lambda} = \begin{bmatrix} P_1 - \frac{E[X_1 Y]^2}{E[Y^2]} & -\frac{E[X_1 Y] E[X_2 Y]}{E[Y^2]} \\ -\frac{E[X_1 Y] E[X_2 Y]}{E[Y^2]} & P_2 - \frac{E[X_2 Y]^2}{E[Y^2]} \end{bmatrix} \tag{13}$$

Using (1), $E[X_1 Y]$, $E[X_2 Y]$ and $E[Y^2]$ can be written in terms of $E[X_1 J]$ and $E[X_2 J]$ as

$$E[X_i Y] = \sqrt{h_i} P_i + \sqrt{\gamma} E[X_i J], \quad i = 1, 2 \tag{14}$$

$$E[Y^2] = h_1 P_1 + h_2 P_2 + 2\sqrt{h_1 \gamma} E[X_1 J] + 2\sqrt{h_2 \gamma} E[X_2 J] + \sigma_N^2 + P_J \tag{15}$$

Therefore, $|\boldsymbol{\Lambda}|$ can be expressed as a function of $E[X_1 J]$ and $E[X_2 J]$. Consider any jamming signal $J$. Define $R$ as

$$R = J - X_1 \frac{E[X_1 J]}{P_1} - X_2 \frac{E[X_2 J]}{P_2} \tag{16}$$

Note that $R$ is uncorrelated with $X_1$ and $X_2$. The power of this jamming signal is

$$E[J^2] = \frac{E[X_1 J]^2}{P_1} + \frac{E[X_2 J]^2}{P_2} + E[R^2] \tag{17}$$

For this jamming signal to be feasible, we should have

$$\frac{E[X_1 J]^2}{P_1} + \frac{E[X_2 J]^2}{P_2} \leq P_J \tag{18}$$

Now define a linear jamming signal as in (6), where $\rho_i = E[X_i J]/P_i$, $i = 1, 2$, and $N_J$ is an independent Gaussian random variable with power $E[R^2]$. This linear jammer has the same power as $J$ and therefore is feasible. Moreover, it results in the same $|\boldsymbol{\Lambda}|$ value as $J$. Hence, for any signal in the set of feasible jamming signals, there is an equivalent linear jamming signal which results in the same upper bound in (12). This means that, there exists a feasible linear jamming signal which is as effective as any other feasible jamming signal, and this concludes the proof.

The next step is to find $\rho_1$ and $\rho_2$ for the linear jamming signal in (6) which achieves the highest upper bound in (12). Since both (11) and (12) hold with equality, the linear jamming parameters that maximize (12), maximize (10), or equivalently, minimize the mutual information in (9). Following the literature [3, 4], we call this mutual information value, the capacity. Using (8)

$$C = \frac{1}{2} \log \left( 1 + \frac{(\sqrt{h_1} + \sqrt{\gamma}\rho_1)^2 P_1 + (\sqrt{h_2} + \sqrt{\gamma}\rho_2)^2 P_2}{\gamma \sigma_{N_J}^2 + \sigma_N^2} \right) \tag{19}$$

which is a monotonically increasing function of the SNR, therefore the jammer's equivalent objective is to minimize the SNR value. We have the following minimization problem

$$\min_{\{\rho_1, \rho_2, \sigma_{N_J}^2\}} \quad \frac{(\sqrt{h_1} + \sqrt{\gamma}\rho_1)^2 P_1 + (\sqrt{h_2} + \sqrt{\gamma}\rho_2)^2 P_2}{\gamma \sigma_{N_J}^2 + \sigma_N^2}$$

$$\text{s.t.} \quad \rho_1^2 P_1 + \rho_2^2 P_2 + \sigma_{N_J}^2 \leq P_J \tag{20}$$

The Karush-Kuhn-Tucker (KKT) necessary conditions are

$$\frac{\sqrt{\gamma}(\sqrt{h_1} + \sqrt{\gamma}\rho_1) P_1}{\gamma \sigma_{N_J}^2 + \sigma_N^2} + \lambda \rho_1 P_1 = 0 \tag{21}$$

$$\frac{\sqrt{\gamma}(\sqrt{h_2} + \sqrt{\gamma}\rho_2) P_2}{\gamma \sigma_{N_J}^2 + \sigma_N^2} + \lambda \rho_2 P_2 = 0 \tag{22}$$

$$-\gamma \frac{(\sqrt{h_1} + \sqrt{\gamma}\rho_1)^2 P_1 + (\sqrt{h_2} + \sqrt{\gamma}\rho_2)^2 P_2}{(\gamma \sigma_{N_J}^2 + \sigma_N^2)^2} + \lambda - \delta = 0 \tag{23}$$

where $\delta$ is the complementary slackness variable for $\sigma_{N_J}^2$. Equations (21) and (22) have the following solution

$$\rho_1 = -\sqrt{h_1} \frac{\gamma P_J + \sigma_N^2}{\sqrt{\gamma}(h_1 P_1 + h_2 P_2)} \tag{24}$$

$$\rho_2 = -\sqrt{h_2} \frac{\gamma P_J + \sigma_N^2}{\sqrt{\gamma}(h_1 P_1 + h_2 P_2)} \tag{25}$$

Therefore whenever these values of $\rho_1$ and $\rho_2$ are feasible, they characterize the best jammer strategy. Ultimately, including the limiting feasible values, the optimum jamming coefficients

are

$$(\rho_1, \rho_2) = \begin{cases} (-\frac{\sqrt{h_1}}{\sqrt{\gamma}}, -\frac{\sqrt{h_2}}{\sqrt{\gamma}}) & \text{if } \gamma P_J \geq h_1 P_1 + h_2 P_2 \\ (-\rho\sqrt{h_1}, -\rho\sqrt{h_2}) & \text{if } \gamma P_J < h_1 P_1 + h_2 P_2 \end{cases} \quad (26)$$

where

$$\rho = \min\left\{\sqrt{\frac{P_J}{h_1 P_1 + h_2 P_2}}, \frac{\gamma P_J + \sigma_N^2}{\sqrt{\gamma}(h_1 P_1 + h_2 P_2)}\right\} \quad (27)$$

and the jammer transmits as in (6). We observe that the amount of power the jammer allocates for jamming each user is proportional to that user's effective received power which is $h_i P_i$ for user $i$, $i = 1, 2$.

Figure 2 shows an example of the jammer decision regions. In region A, $(\rho_1, \rho_2) = (-\sqrt{h_1}/\sqrt{\gamma}, -\sqrt{h_2}/\sqrt{\gamma})$ and the jammer only uses enough power to zero out the transmitted signals. In region B, $(\rho_1, \rho_2) = (-\sqrt{h_1}, -\sqrt{h_2})\sqrt{P_J/(h_1 P_1 + h_2 P_2)}$ and the jammer uses all of its power to cancel the transmitted signals as much as possible. In region C, $(\rho_1, \rho_2) = (-\sqrt{h_1}, -\sqrt{h_2})(\gamma P_J + \sigma_N^2)/(\sqrt{\gamma}(h_1 P_1 + h_2 P_2))$ and the jammer uses part of its power to cancel the transmitted signals, and the rest of its power to add Gaussian noise to the transmitted signal. Therefore, for low channel coefficients where the effective received powers of the users are small, the optimum jamming strategy is to subtract the user signals as much as possible, while in high channel coefficients, the jammer uses its power both for adding Gaussian noise and for correlating with the user signals.

## 3.2 Jamming with Eavesdropping Information

Now suppose that the jammer gains information about the user signals only through an AWGN eavesdropping channel,

$$Y_e = \sqrt{g_1} X_1 + \sqrt{g_2} X_2 + N_e \quad (28)$$

We define linear jamming as transmitting a linear combination of the signal received at the jammer $Y_e$ and Gaussian noise, i.e.,

$$J = \rho Y_e + N_J \quad (29)$$

Here, we will prove that in the eavesdropping case as well, linear jamming and Gaussian signalling is a game solution. The proof of the optimality of Gaussian signalling, when the jammer is linear is similar to the previous section as follows. Using (1), (28) and (29), if the jammer is linear, the received signal is

$$Y = (\sqrt{h_1} + \rho\sqrt{\gamma g_1})X_1 + (\sqrt{h_2} + \rho\sqrt{\gamma g_2})X_2 + \rho\sqrt{\gamma}N_e + \sqrt{\gamma}N_J \qquad (30)$$

which is an AWGN multiple access channel, therefore, the best signalling for the users is Gaussian.

However, when it comes to showing the optimality of linear jamming when the users employ Gaussian signalling, the method of the previous section cannot be used, since from (28) and (29), the values of $E[X_1 J]$ and $E[X_2 J]$ that are achievable through linear jamming, should further satisfy

$$\frac{E[X_1 J]}{\sqrt{g_1}} = \frac{E[X_2 J]}{\sqrt{g_2}} \qquad (31)$$

Therefore, linear jamming may not achieve all $|\Lambda|$ values in (12) that are allowed under the power constraints. Here, we show the optimality of linear jamming, by setting up an equivalent multiple access channel. Define random variables $Z_1$ and $Z_2$ in terms of $X_1$ and $X_2$ as

$$Z_1 = X_1 + \frac{\sqrt{g_2}}{\sqrt{g_1}} X_2 \qquad (32)$$

$$Z_2 = -\frac{\sqrt{g_1 g_2}P_2}{g_1 P_1 + g_2 P_2} X_1 + \frac{g_1 P_1}{g_1 P_1 + g_2 P_2} X_2 \qquad (33)$$

It is straightforward to verify that $Z_1$ and $Z_2$ are uncorrelated, and hence, independent Gaussian random variables. Moreover, since the two pairs have a one-to-one relation, they result in the same input/output mutual information, i.e., $I(X_1, X_2; Y) = I(Z_1, Z_2; Y)$ [7]. Therefore, the game's objective function can be replaced with $I(Z_1, Z_2; Y)$. Now, using (1), (32) and (33), we can rewrite $Y$ in terms of $Z_1$ and $Z_2$ as

$$Y = u_1 Z_1 + u_2 Z_2 + \sqrt{\gamma}J + N \qquad (34)$$

where

$$u_1 = \frac{\sqrt{g_1}}{g_1 P_1 + g_2 P_2}(\sqrt{g_1 h_1}P_1 + \sqrt{g_2 h_2}P_2) \qquad (35)$$

$$u_2 = \sqrt{h_2} - \sqrt{h_1}\frac{\sqrt{g_2}}{\sqrt{g_1}} \qquad (36)$$

We can also write the eavesdropping signal received at the jammer using (28), (32) and (33)

as
$$Y_e = \sqrt{g_1}Z_1 + N_e \tag{37}$$

Note that $Y_e$ is independent of $Z_2$. Equations (34) and (37) define a two user, one jammer system, depicted in Figure 3, where the jammer has eavesdropping information only about one of the users, which is the key in proving the optimality of linear jamming as follows. We rewrite the equivalent input/output mutual information as

$$I(Z_1, Z_2; Y) = h(Z_1, Z_2) - h(Z_1, Z_2|Y) \tag{38}$$

The jammer's strategy can only affect the second term above,

$$h(Z_1, Z_2|Y) = h(Z_1 - a_1Y, Z_2 - a_2Y|Y) \tag{39}$$
$$\leq h(Z_1 - a_1Y, Z_2 - a_2Y) \tag{40}$$
$$\leq \frac{1}{2}\log(1 + |\boldsymbol{\Sigma}|) \tag{41}$$

where $\boldsymbol{\Sigma}$ is the covariance matrix of $Z_1 - a_1Y$ and $Z_2 - a_2Y$. Following steps similar to those in the previous section, when the users are Gaussian, employing linear jamming together with a good choice of $a_1$ and $a_2$ can make both inequalities hold with equality, and $|\boldsymbol{\Sigma}|$ will only be a function of $E[Z_1 J]$ and $E[Z_2 J]$. However, $Z_2$ and $J$ are independent and $E[Z_2 J] = 0$, therefore, $|\boldsymbol{\Sigma}|$ is only a function of $E[Z_1 J]$. In the sequel, we show that all $E[Z_1 J]$ values that are achievable by all feasible jamming signals, are also achievable by some feasible linear jamming signal, and therefore, linear jamming achieves (40) and (41) with equality and also achieves the largest possible upper bound in (41).

Using (37), the linear least squared error (LLSE) estimate of $Z_1$ from $Y_e$ is [11]

$$\tilde{Z}_1(Y_e) = \frac{E[Z_1 Y_e]}{E[Y_e^2]} Y_e \tag{42}$$
$$= \frac{\sqrt{g_1}E[Z_1^2]}{\sigma_{N_e}^2 + g_1 E[Z_1^2]} Y_e \tag{43}$$

and the LLSE estimate error is

$$E[(\tilde{Z}_1(Y_e) - Z_1)^2] = \frac{\sigma_{N_e}^2 E[Z_1^2]}{\sigma_{N_e}^2 + g_1 E[Z_1^2]} \tag{44}$$

Since $Z_1$ and $N_e$ are Gaussian, this estimate is also the minimum mean squared error (MMSE) estimate of $Z_1$, therefore, any other estimate of $Z_1$ results in a higher mean squared error. Now consider any jamming signal $J$ which is a function of $Y_e$, i.e., $J = f(Y_e)$, where $f$ is a potentially random function. The LLSE estimate of $Z_1$ from $J$ is

$$\hat{Z}_1 = \frac{E[Z_1 J]}{E[J^2]} J \qquad (45)$$

and the estimate error is

$$E[(\hat{Z}_1 - Z_1)^2] = E[Z_1^2] - \frac{E^2[Z_1 J]}{E[J^2]} \qquad (46)$$

This is also another estimator of $Z_1$ from $Y_e$, hence the estimation error in $\hat{Z}_1$ is greater than or equal to the estimation error in $\tilde{Z}_1$

$$E[Z_1^2] - \frac{E^2[Z_1 J]}{E[J^2]} \geq \frac{\sigma_{N_e}^2 E[Z_1^2]}{\sigma_{N_e}^2 + g_1 E[Z_1^2]} \qquad (47)$$

Therefore, the feasible values of $E[Z_1 J]$ should satisfy

$$E^2[Z_1 J] \leq \frac{g_1 E^2[Z_1^2] E[J^2]}{\sigma_{N_e}^2 + g_1 E[Z_1^2]} \qquad (48)$$

$$= \frac{g_1 E^2[Z_1^2] P_J}{\sigma_{N_e}^2 + g_1 E[Z_1^2]} \qquad (49)$$

Meanwhile, using (37), the achievable $\rho$ values for a linear jammer satisfy

$$\rho^2 \leq \frac{P_J}{\sigma_{N_e}^2 + g_1 E[Z_1^2]} \qquad (50)$$

Also, from (29) and (37), for a linear jammer, $E[Z_1 J] = \rho \sqrt{g_1} E[Z_1^2]$, which together with (50) results in that linear jamming can achieve all $E[Z_1 J]$ values satisfying

$$E^2[Z_1 J] \leq \frac{g_1 E^2[Z_1^2] P_J}{\sigma_{N_e}^2 + g_1 E[Z_1^2]} \qquad (51)$$

The right hand sides of (49) and (51) are identical, where the former limits the $E[Z_1 J]$ values for all feasible jammers, and the latter describes all the $E[Z_1 J]$ values that are achievable with linear jamming. We conclude that for any signal in the set of feasible jamming signals, there is an equivalent linear jamming signal which results in the same upper bound in (41), which means that there exists a feasible linear jamming signal which is as effective as any other feasible jamming signal, and this concludes the proof.

We now derive the jamming coefficient for an optimal linear jammer with eavesdropping information. Using (28) and (29), the jamming signal is

$$J = \rho(\sqrt{g_1}X_1 + \sqrt{g_2}X_2 + N_e) + N_J \tag{52}$$

and the received signal is as in (30). The jammer's optimization problem is

$$\min_{\{\rho,\sigma_{N_J}^2\}} \frac{(\sqrt{h_1} + \rho\sqrt{\gamma g_1})^2 P_1 + (\sqrt{h_2} + \rho\sqrt{\gamma g_2})^2 P_2}{\gamma\rho^2\sigma_{N_e}^2 + \gamma\sigma_{N_J}^2 + \sigma_N^2}$$
$$\text{s.t.} \quad \rho^2(g_1 P_1 + g_2 P_2 + \sigma_{N_e}^2) + \sigma_{N_J}^2 \leq P_J \tag{53}$$

The KKTs for this problem result in a third degree equation in $\rho$ and can be solved using numerical optimization.

Figure 4 shows the SNR as a function of one of the channel coefficients $h_1$. The SNR is compared in two scenarios, when the jammer eavesdrops, and when it has full information about the user signals. We observe that at very low $h_1$, the patterns of the two scenarios differ considerably, while for very large values of $h_1$, they follow the same monotone SNR pattern. This is in fact expected, since at very small channel attenuations, the jammer with complete information is able to cancel a good portion of the user signals, while the noise in the eavesdropping channel restricts the eavesdropping jammer in doing the same. Also, when the channel attenuation is very high, in both scenarios, the jammer uses most of its power for adding noise, and therefore, they both follow the same pattern. However, when the jammer has full information about the user signals, the jamming coefficients are proportional to the user channel attenuations, and therefore, the jamming coefficient for the second user is very small compared to the first user, while when the jammer has eavesdropping information, the jamming coefficient is the same for both users. This causes the difference between the two scenarios at high SNR.

## 4 Jamming in Fading Multi-user AWGN Channels

We now investigate the optimum user/jammer strategies when the channels are fading. Throughout this section, we use the term CSI, for the channel state information on the links from the users to the receiver, and assume that the link between the jammer and the receiver is non-fading. This section is divided into three parts corresponding to three different assumptions: 1) no CSI at the transmitters, 2) uncorrelated jamming with full CSI at the transmitters, and 3) correlated jamming with full CSI at the transmitters. In each part, the receiver is assumed to know the CSI, while various assumptions are made on the availability of the CSI at the jammer.

The problem of correlated jamming in single-user MIMO fading channels, with the assumption that the transmitter and the jammer do not have the CSI, but the receiver has the CSI, has been investigated in [4], and it has been shown that the best strategies for the user and the jammer is to evenly spread their powers over their corresponding transmit antennas, and transmit independent Gaussian signals, and the jammer is better off disregarding its information about the user signal. The problem of uncorrelated jamming in single-user MISO fading channels has also been investigated in [5], where the user and jammer are restricted to employ Gaussian signalling. In [5], both the user channel and the jammer channel are considered to have fading, and also it is assumed that the user and jammer may or may not have access to the CSI of their own channels, but they do not have access to the CSI of their opponent's channel.

## 4.1 No CSI at the Transmitters

When the transmitters do not have the CSI, it is reasonable to assume that the jammer does not have the CSI either. In the sequel, we show that the jammer's information about the transmitted signals will be irrelevant and therefore, it will not make any difference whether it has perfect or noisy information about the transmitted signals. This is a multi-user generalization of the results of [4] in a SISO system.

Assuming that the user links are fading and the jammer link is non-fading, the received signal is

$$Y = H_1 X_1 + H_2 X_2 + \sqrt{\gamma} J + N \tag{54}$$

The receiver is assumed to know the fading coefficients while the users and the jammer only know the fading statistics. Here, we assume that all the random variables are complex valued and $H_1$ and $H_2$ are circularly symmetric complex Gaussian. Following [4] in finding the Nash equilibrium solution of the mutual information game by conditioning on the fading coefficients,

$$I(Y, H_1, H_2; X_1, X_2) = h(X_1, X_2) - h(X_1, X_2 | Y, H_1, H_2) \tag{55}$$
$$= h(X_1, X_2) - h(X_1 - A_1 Y, X_2 - A_2 Y | Y, H_1, H_2) \tag{56}$$

where $A_1$ and $A_2$ are functions of $H_1$ and $H_2$. The jammer's strategy can only affect the second term above

$$h(X_1 - A_1 Y, X_2 - A_2 Y | Y, H_1, H_2) \leq h(X_1 - A_1 Y, X_2 - A_2 Y | H_1, H_2) \tag{57}$$
$$\leq E\left[\frac{1}{2} \log\left((2\pi e)^2 |\mathbf{\Lambda}|\right)\right] \tag{58}$$

where $\mathbf{\Lambda}$ is the covariance matrix of

$$(X_1 - A_1Y, X_2 - A_2Y | H_1, H_2)$$

and the expectation in (58) is over the joint distribution of $H_1$ and $H_2$. Choosing

$$A_i = \frac{E[X_iY|H_1, H_2]}{E[Y^2|H_1, H_2]}, \quad i = 1, 2 \tag{59}$$

makes $X_1 - A_1Y$ and $X_2 - A_2Y$ conditionally uncorrelated with $Y$, given $H_1$ and $H_2$. Following arguments similar to those in the previous section, since $X_1$ and $X_2$ are Gaussian, linear jamming makes $J$, and therefore $Y$ jointly Gaussian with $X_1$ and $X_2$, and the above two inequalities hold with equality. Now, we need to show that linear jamming can also achieve the highest upper bound in the second inequality. First note that $X_1$, $X_2$ and $J$ are independent of $H_1$ and $H_2$, hence

$$E[X_iY|H_1, H_2] = H_iP_i + \sqrt{\gamma}E[X_iJ|H_1, H_2] \tag{60}$$
$$= H_iP_i + \sqrt{\gamma}E[X_iJ], \quad i = 1, 2 \tag{61}$$

where (61) holds since $X_1$, $X_2$ and $J$ are independent of $H_1$ and $H_2$. Therefore, the second upper bound above is only a function of $E[X_iJ]$, $i = 1, 2$. The rest of the arguments in the previous section follow, resulting in that linear jamming can achieve the highest upper bound in the second inequality, which concludes the proof.

The strategies corresponding to the game solution will be Gaussian signalling and linear jamming. The jamming signal is as in (6), and the received signal is

$$Y = (H_1 + \sqrt{\gamma}\rho_1)X_1 + (H_2 + \sqrt{\gamma}\rho_2)X_2 + \sqrt{\gamma}N_J + N \tag{62}$$

The last step is to find the best $\rho_1$ and $\rho_2$. Given that the jammer knows the statistics of the fading, its optimization problem is

$$\min_{\{\rho_1, \rho_2, \sigma_{N_J}^2\}} \frac{1}{2}E\left[\log\left(1 + \frac{|H_1 + \sqrt{\gamma}\rho_1|^2P_1 + |H_2 + \sqrt{\gamma}\rho_2|^2P_2}{\gamma\sigma_{N_J}^2 + \sigma_N^2}\right)\right] \tag{63}$$

$$\text{s.t.} \quad \rho_1^2P_1 + \rho_2^2P_2 + \sigma_{N_J}^2 \leq P_J \tag{64}$$

The function in (63) is very similar to (19), except for the expectation taken over the channel states. The jamming coefficients $\rho_1$ and $\rho_2$ in (63) and (64) are independent of $H_1$ and $H_2$. Distributions of $H_1$ and $H_2$ are centered around zero, therefore intuitively, shifting them will make their norms larger. This fact can also be derived using [12, Theorem 1]. Therefore, the optimum jamming coefficients are $\rho_1 = \rho_2 = 0$, and the jammer disregards its complete information. We conclude that if the jammer's information is noisy, it cannot do any better than what it did when it had noiseless information, and therefore, it should disregard the incomplete information, whether the incompleteness is because of fading or AWGN or both in the jammer's eavesdropping channel.

## 4.2 Uncorrelated Jamming with CSI at the Transmitters

We now consider a two user fading channel with a jammer who does not have any information about the user signals and therefore, is uncorrelated with the users. We also assume that the user links are fading and the state of the user links are known to the users. The users are now able to distribute their powers optimally over the user channel states. Capacity of fading channels with CSI both at the transmitter and the receiver when there is no jammer, has been investigated in [13] and [14], and optimum signalling and power allocation strategies have been derived. In this section, we consider the same problem when there is a jammer in the system. We first consider the single-user case and assume that the jamming channel is non-fading

$$Y = HX + \sqrt{\gamma}J + N \tag{65}$$

When the CSI is available both at the transmitter and the receiver, the maximum input/output mutual information is

$$C = I(X;Y|H) \tag{66}$$

where $X$ is a random variable whose conditional distribution conditioned on $H$, is chosen to maximize $C$. The conditional input/output mutual information $I(X;Y|H)$ is a convex function of $f(y|x,h)$ for any fixed conditional input distribution $f(x|h)$, and a concave function of $f(x|h)$ for any fixed conditional transition distribution $f(y|x,h)$. At each channel state, there is a saddle point which is to employ Gaussian signalling and linear jamming. This specifies the solution to the mutual information sub-game at any given channel state. Moreover, if a saddle point exists over all possible power allocation strategies of the user and the jammer, under user and jammer power constraints, that saddle point power allocation along with the signalling and jamming strategies specified as the solution of the sub-games corresponding to each channel state, will give the overall solution.

We proceed with first assuming that even though the users have the CSI, the jammer does not have the CSI, and then assuming that the CSI is available both at the users and at the jammer. The latter is a reasonable assumption, since we can assume that the jammer eavesdrops the communication link from the receiver to the transmitter, where the receiver sends the CSI feedback information to the user.

If the jammer has no information about the fading channel state, the best strategy for the jammer is to transmit Gaussian noise. The received signal at fading level $h$ is

$$Y = \sqrt{h}X + \sqrt{\gamma}N_J + N \qquad (67)$$

and the capacity is

$$C = \frac{1}{2}E\left[\log\left(1 + \frac{hP(h)}{\sigma_N^2 + \gamma P_J}\right)\right] \qquad (68)$$

where $P(h)$ is the user power at fading level $h$ which should satisfy

$$E[P(h)] \leq P \qquad (69)$$

The best user power allocation is waterfilling over the equivalent parallel AWGN channels [13] with equivalent noise levels $(\sigma_N^2 + \gamma P_J)/h$, i.e.,

$$P(h) = \left(\frac{1}{\lambda} - \frac{\sigma_N^2 + \gamma P_J}{h}\right)^+ \qquad (70)$$

where $(x)^+ = \max(x, 0)$, and $\lambda$ is a constant chosen to enforce the user power constraint. The corresponding two user system, where the jammer is not aware of the user channel coefficients, is a straightforward extension of the results in [15] where only one user transmits at a time. The jammer will again use all its power to add Gaussian noise.

Next, we assume that the uncorrelated jammer has the CSI as well. The received signal in the single-user system is the same as in (67). At each channel state, the jammer transmits Gaussian noise at the power level allocated to that state. The capacity is

$$C = \frac{1}{2}E\left[\log\left(1 + \frac{hP(h)}{\sigma_N^2 + \gamma J(h)}\right)\right] \qquad (71)$$

where $J(h)$ is the jammer power at fading level $h$. The user power constraint is the same as (69), and the jammer power constraint is

$$E[J(h)] \leq P_J \qquad (72)$$

Since every term of the capacity corresponding to a channel state $h$ is concave in $P(h)$ for fixed $J(h)$ and convex in $J(h)$ for fixed $P(h)$, the capacity is a concave function of $P$ for fixed $J$ and a convex function of $J$ for fixed $P$. Given the convexity/concavity properties of the capacity and using [9, Theorem 16, p. 75], [10, Proposition 2.6.9], the set of saddle points is compact and nonempty and therefore the mutual information game has a solution. At the game solution, the pair of strategies should satisfy the KKTs of the two optimization problems corresponding to the user and the jammer. The user maximizes (71) subject to (69), while the jammer minimizes (71) subject to (72). Writing the KKTs for each state-allocated user power, we get

$$-\frac{h}{\sigma_N^2 + \gamma J(h) + hP(h)} + \lambda - \xi(h) = 0 \qquad (73)$$

where $\xi(h)$ is the complementary slackness variable for $P(h)$. Similarly, writing the KKTs for each state-allocated jammer power, we get

$$-\frac{\gamma h P(h)}{(\sigma_N^2 + \gamma J(h))(\sigma_N^2 + \gamma J(h) + hP(h))} + \mu - \delta(h) = 0 \qquad (74)$$

where $\delta(h)$ is the complementary slackness variable for $J(h)$.

The optimum strategies should solve (73) and (74) simultaneously. There are four possible cases at each fading level. Case 1: $P(h) > 0$ and $J(h) > 0$, case 2: $P(h) = 0$ and $J(h) > 0$, case 3: $P(h) > 0$ and $J(h) = 0$ and case 4: $P(h) = 0$ and $J(h) = 0$. If $P(h) = 0$, (74) cannot be satisfied unless $\delta(h) > 0$, therefore, case 2 never happens. This is expected, because if the user does not transmit at a fading level, the jammer does not gain anything by transmitting at that fading level. In case 1, both complementary slackness variables are zero, and (73) and (74) become

$$\frac{h}{\sigma_N^2 + \gamma J(h) + hP(h)} = \lambda \qquad (75)$$

$$\frac{\gamma h P(h)}{(\sigma_N^2 + \gamma J(h))(\sigma_N^2 + \gamma J(h) + hP(h))} = \mu \qquad (76)$$

which result in a linear relation between the user and jammer power allocations

$$\frac{\sigma_N^2 + \gamma J(h)}{\gamma P(h)} = \frac{\lambda}{\mu} \qquad (77)$$

and solving for the user's power

$$P(h) = \frac{h}{\lambda(h + \gamma \frac{\lambda}{\mu})} \qquad (78)$$

which is a monotonically increasing function of the fading variable $h$. Therefore, at any fading level where the user and jammer powers are nonzero, the jammer's power is a linear function of the user's, and they both allocate more power to better channel states. Case 1 is valid as long as (77) and (78) result in positive $J(h)$, which is for $P(h) > \sigma_N^2 \mu/(\gamma \lambda)$ or $h > \sigma_N^2 \gamma \lambda/(\gamma - \sigma_N^2 \mu)$. For $h < \sigma_N^2 \gamma \lambda/(\gamma - \sigma_N^2 \mu)$, $J(h) = 0$ which results in cases 3 and 4 combined. In this case (73) will turn to waterfilling. Therefore, combining all of these, the optimum power allocations are

$$P(h) = \begin{cases} (\frac{1}{\lambda} - \frac{\gamma \sigma_N^2}{h})^+ & \text{if } h < \frac{\sigma_N^2 \gamma \lambda}{\gamma - \sigma_N^2 \mu} \\ \frac{h}{\lambda(h + \gamma \frac{\lambda}{\mu})} & \text{if } h \geq \frac{\sigma_N^2 \gamma \lambda}{\gamma - \sigma_N^2 \mu} \end{cases} \quad (79)$$

and

$$J(h) = \begin{cases} 0 & \text{if } h < \frac{\sigma_N^2 \gamma \lambda}{\gamma - \sigma_N^2 \mu} \\ \frac{h}{\mu(h + \gamma \frac{\lambda}{\mu})} - \frac{\sigma_N^2}{\gamma} & \text{if } h \geq \frac{\sigma_N^2 \gamma \lambda}{\gamma - \sigma_N^2 \mu} \end{cases} \quad (80)$$

where $\lambda$ and $\mu$ are found using the user and jammer power constraints.

Figure 5 shows $P(h)$ and $J(h)$ for a simple example, where the fading is assumed to be Rayleigh with parameter 1. Figure 5 also includes the power allocation curve for a case where there is no jammer [13], which, compared to the case with a jammer, shows that the presence of the jammer changes the power allocation strategy of the user. When there is a jammer, from our closed form solutions in (79) and (80), and from Figure 5, we observe that both the user and the jammer keep quiet at very low fading levels. Then, as the user channel gets better, the user starts transmitting, with more power allocated to better channels, and eventually at even better channels, the jammer starts jamming, again with more power allocated to better channels.

We now discuss the two user system where the jammer is uncorrelated but it has access to CSI. The power allocation strategies will be functions of the two channel states $\mathbf{h} = (h_1, h_2)$. The capacity is

$$C = \frac{1}{2} E \left[ \log \left( 1 + \frac{h_1 P_1(\mathbf{h}) + h_2 P_2(\mathbf{h})}{\sigma_N^2 + J(\mathbf{h})} \right) \right] \quad (81)$$

The KKTs for the users result in

$$-\frac{h_i}{\sigma_N^2 + J(\mathbf{h}) + h_1 P_1(\mathbf{h}) + h_2 P_2(\mathbf{h})} + \lambda_i - \gamma_i(\mathbf{h}) = 0, \quad i = 1, 2 \quad (82)$$

where $\gamma_i(\mathbf{h})$ is the complementary slackness variables for $P_i(\mathbf{h})$, $i = 1, 2$. If at a pair of fading levels, both users transmit with nonzero powers, (82) results in

$$\frac{h_1}{h_2} = \frac{\lambda_1}{\lambda_2} \quad (83)$$

which happens with zero probability if the fading PDF is continuous. Therefore, similar to the system without a jammer in [15], only one user transmits at any given channel state.

Define $h$ as

$$h = \max\left(\frac{h_1}{\lambda_1}, \frac{h_2}{\lambda_2}\right) \tag{84}$$

Now, the users and the jammer power allocations are functions of $h$, therefore we can replace **h** by $h$ in the previous equations. Since the users do not transmit at the same time, we can use the single-user results to find the user power allocations

$$P_i(h) = \begin{cases} 0 & \text{if } h \neq \frac{h_i}{\lambda_i} \\ \frac{1}{h_i} q(h) & \text{if } h = \frac{h_i}{\lambda_i} \end{cases}, \quad i = 1, 2 \tag{85}$$

where $q(h)$ is

$$q(h) = \begin{cases} (h - \sigma_N^2)^+ & \text{if } h < \frac{\sigma_N^2}{1 - \sigma_N^2 \mu} \\ \frac{h}{1 + \frac{1}{h\mu}} & \text{if } h \geq \frac{\sigma_N^2}{1 - \sigma_N^2 \mu} \end{cases} \tag{86}$$

The jammer's power allocation is

$$J(h) = \begin{cases} 0 & \text{if } h < \frac{\sigma_N^2}{1 - \sigma_N^2 \mu} \\ \frac{1}{\mu(1 + \frac{1}{h\mu})} - \sigma_N^2 & \text{if } h \geq \frac{\sigma_N^2}{1 - \sigma_N^2 \mu} \end{cases} \tag{87}$$

The strategies follow a pattern as in Figure 6, that is, the users do not transmit simultaneously, no party transmits at very low fading levels, as the channels get better, the user with a relatively better channel transmits, and eventually the jammer starts transmitting at even better channels. The threshold values $u_1$, $u_2$, $v_1$ and $v_2$ are to be chosen to satisfy the power constraints.

## 4.3 Correlated Jamming with CSI at the Transmitters

In this section, we consider a two user fading channel with a jammer who knows the user signals and therefore, is correlated with the users. We assume that the user links are fading and the state of the user links are known to the users and the correlated jammer. The users and the jammer are now able to distribute their powers optimally over the user channel states. The jammer link is again assumed to be non-fading. We first show that this game does not always have a Nash equilibrium solution, and then, we find the max-min user strategies and the corresponding jamming strategy. The max-min user power allocation corresponds to the users' best power allocation, in a situation where the user chooses its strategy only once, while after the user chooses its strategy, the jammer can observe it and choose the corresponding best jamming strategy. Note that if the game had a solution, max-min and min-max strategies would have been the same, and would also be the same as the game solution.

As in the previous sections we start with a single-user system. The input/output mutual information is as in (66), which is a weighted sum of the input/output mutual information at each channel state. The user and jammer power constraints can be written as

$$E\left[E[X^2|H]\right] \leq P \tag{88}$$

$$E\left[E[J^2|H]\right] \leq P_J \tag{89}$$

$E[X^2|H = h]$ and $E[J^2|H = h]$ are the user and jammer powers allocated to the fading level $H = h$. Any pair of user and jammer strategies, results in a pair of user and jammer power allocation strategies over the user channel fading distribution. Therefore, the game's solution can be described as a pair of user and jammer power allocation strategies, along with the user and jammer signalling functions at each channel state. Using our results for the non-fading channels, we have that irrespective of the existence or non-existence of optimal power allocation functions for the user and the jammer, the sub-games always have a Nash equilibrium solution at each channel state, under any pair of user and jammer power allocation functions. The solution of the sub-games at each channel state is Gaussian signalling for the users and linear jamming for the jammer.

Since the jammer knows the channel state and the transmitted signal, the received signal is

$$Y = \left(\sqrt{h} + \rho(h)\right) X + N_J + N \tag{90}$$

where the variance of $N_J$ is also a function of $h$, $\sigma^2_{N_J}(h)$. Given a pair of power allocation functions $P(h)$ and $J(h)$, the capacity is

$$C = \frac{1}{2} E\left[\log\left(1 + \frac{\left(\sqrt{h} + \rho(h)\right)^2 P(h)}{\sigma^2_N + \sigma^2_{N_J}(h)}\right)\right] \tag{91}$$

where for each channel state $h$, $\rho(h)$ and $\sigma^2_{N_J}(h)$ are the optimal linear jamming coefficients for an equivalent non-fading channel as given in Section 3.1, with attenuation $h$ and user and jammer powers $P(h)$ and $J(h)$. The power constraints of the user and the jammer are as in (69) and (72).

The capacity here does not have the convexity/concavity properties in the user and jammer power allocation functions. In the sequel, we show that a pair of strategies, which is simultaneously optimal for the user and the jammer, does not always exist. We first assume that the user chooses its strategy once at the beginning of the communication, knowing that the jammer will employ the corresponding optimal jamming strategy. We then characterize the user and jammer strategies in this scenario. If the game had a Nash equilibrium solution, it would have been this pair of user and jammer strategies, however we prove the converse. We consider the resulting jamming strategy and assume that the jammer chooses this strategy at the beginning of the communication, and show that if the user had the possibility of changing its strategy, the current user strategy would have not been optimal.

First, given any user power allocation function $P(h)$, we find the jammer's best response, where a best response describes what a player does, given the other player's move [8]. In this case, a best response is the jammer's best power allocation strategy, given a fixed user power allocation strategy. The jammer's best response can also be thought of as a pair of functions $\rho(h)$ and $\sigma_{N_J}^2(h)$ which minimizes the capacity

$$C = \frac{1}{2} E \left[ \log \left( 1 + \frac{\left(\sqrt{h} + \rho(h)\right)^2 P(h)}{\sigma_N^2 + \sigma_{N_J}^2(h)} \right) \right] \tag{92}$$

and $\rho(h)$ and $\sigma_{N_J}^2(h)$ are constrained such that

$$E\left[\rho^2(h)P(h) + \sigma_{N_J}^2(h)\right] \leq P_J \tag{93}$$

and $\sigma_{N_J}^2(h)$ is nonnegative. The first order KKT conditions for the jammer are

$$\frac{\sqrt{h} + \rho(h)}{\sigma_N^2 + \sigma_{N_J}^2(h) + (\sqrt{h} + \rho(h))^2 P(h)} + \lambda \rho(h) = 0 \tag{94}$$

$$-\frac{(\sqrt{h} + \rho(h))^2 P(h)}{(\sigma_N^2 + \sigma_{N_J}^2(h))(\sigma_N^2 + \sigma_{N_J}^2(h) + (\sqrt{h} + \rho(h))^2 P(h))} + \lambda + \xi(h) = 0 \tag{95}$$

where $\xi(h)$ is the complementary slackness variable for $\sigma_{N_J}^2(h)$. Whenever $\xi(h) > 0$, the jammer uses all its power to correlate with the user signal, and the optimum jamming coefficient should satisfy

$$\frac{\sqrt{h} + \rho(h)}{\sigma_N^2 + (\sqrt{h} + \rho(h))^2 P(h)} + \lambda \rho(h) = 0 \tag{96}$$

which does not result in a closed form solution for the jammer best response. However, in order to derive the max-min user strategy, we need to have the jamming best response in terms of the user power allocation. To make the problem tractable, assume that $\sigma_N^2 = 0$. Now at any channel state that the user transmits, the jammer should also transmit, and the optimal jamming strategy should be such that $\sigma_{N_J}^2(h) > 0$, since otherwise, the capacity would be infinite. The KKTs result in

$$\rho(h) = -\min\left(\frac{1}{\lambda\sqrt{h}P(h)}, \sqrt{h}\right) \tag{97}$$

$$\sigma_{N_J}^2(h) = \left(\frac{1}{\lambda} - \frac{1}{\lambda^2 h P(h)}\right)^+ \tag{98}$$

The optimal jamming strategy is

$$(\rho(h), \sigma_{N_J}^2(h)) = \begin{cases} (-\sqrt{h}, 0) & \text{if } hP(h) \leq \frac{1}{\lambda} \\ \left(-\frac{1}{\lambda\sqrt{h}P(h)}, \frac{1}{\lambda} - \frac{1}{\lambda^2 h P(h)}\right) & \text{if } hP(h) > \frac{1}{\lambda} \end{cases} \tag{99}$$

where $\lambda$ is chosen to satisfy the jammer's power constraint. The total power that the jammer allocates to each channel state is found as

$$J(h) = \rho^2(h)P(h) + \sigma_{N_J}^2(h)$$
$$= \begin{cases} hP(h) & \text{if } hP(h) \leq \frac{1}{\lambda} \\ \frac{1}{\lambda} & \text{if } hP(h) > \frac{1}{\lambda} \end{cases} \tag{100}$$

which describes the best response of the jammer, to the user power allocation $P(h)$, and is shown in Figure 7. Note that Figure 7 shows the best response jammer power allocation as a function of $hP(h)$ and not $P(h)$. The capacity can now be written as a function of the user power allocation alone

$$C = \frac{1}{2}E\left[\log\left(1 + \frac{\left(\sqrt{h} - \min\left(\frac{1}{\lambda\sqrt{h}P(h)}, \sqrt{h}\right)\right)^2 P(h)}{\left(\frac{1}{\lambda} - \frac{1}{\lambda^2 h P(h)}\right)}\right)\right] \tag{101}$$

We now derive the best user power allocation that maximizes this capacity. First note that the function inside the expectation in (101) is zero for $hP(h) \leq 1/\lambda$, therefore, in the optimal user power allocation, $P(h)$ is either zero, or such that $hP(h) > 1/\lambda$. The capacity can now be written as

$$C = \frac{1}{2}E\left[\log\left(1 + \frac{\left(\sqrt{h} - \frac{1}{\lambda\sqrt{h}P(h)}\right)^2 P(h)}{\left(\frac{1}{\lambda} - \frac{1}{\lambda^2 h P(h)}\right)}\right)\right] \tag{102}$$

$$= \frac{1}{2}E\left[\log\left(\lambda\sqrt{h}P(h)\right)\right] \tag{103}$$

The KKT condition for the user power, whenever the user transmits, results in $P(h) = \frac{1}{\mu}$,

for which the total power that the jammer allocates to the channel states is found as

$$J(h) = \begin{cases} 0 & \text{if } h \leq \frac{\mu}{\lambda} \\ \frac{1}{\lambda} & \text{if } h > \frac{\mu}{\lambda} \end{cases} \qquad (104)$$

The user max-min power allocation and the corresponding jammer power allocations are illustrated in Figure 8 where $\mu$ and $\lambda$ are chosen to satisfy the user and jammer power constraints.

Now, we show that the pair of user and jammer power allocations corresponding to the user's max-min solution does not correspond to the Nash equilibrium solution of the game. We consider the jamming strategy in (104) and assume that the jammer chooses this strategy at the beginning of the communication, and show that the current user strategy is not optimal. Consider two fading levels $u^+ > \mu/\lambda$ and $u^- < \mu/\lambda$ in the vicinity of $h = \mu/\lambda$ and close enough to $h = \mu/\lambda$ such that

$$u^- \simeq u^+ \simeq \mu/\lambda \qquad (105)$$

We have $J(u^-) = 0$ and $J(u^+) = 1/\lambda$, hence, $u^-$ and $u^+$ correspond to two channel states which are almost identical in their fading levels, while the jammer is active only in $u^+$. Obviously, it is not optimal for the user to transmit at $u^+$ while not transmitting at $u^-$, therefore, the pair of the user and jammer power allocations derived (which is the user max-min solution), is not a game solution, and the game does not admit a solution.

Even though the max-min solution derived above is not the game Nash equilibrium solution, it is the optimal pair of user and jammer strategies in a system where a conservative user would like to guarantee itself with some capacity value. It also describes the best strategy for a user which is less dynamic than the jammer in terms of changing the transmission strategy, and can choose its strategy only once.

We now extend the max-min results in the single-user system to a two user system. We show that if $h_1$ and $h_2$ are the fading levels of the first and the second user channels, again only one user transmits at any given $\mathbf{h} = (h_1, h_2)$. First, given any pair of user power allocation functions $P_1(\mathbf{h})$ and $P_2(\mathbf{h})$, we find the jammer's best response power allocation strategy. The jammer's best response can also be thought of as a set of functions $\rho_1(\mathbf{h}), \rho_2(\mathbf{h})$ and $\sigma_{N_J}^2(\mathbf{h})$ which minimizes the capacity

$$C = \frac{1}{2}E\left[\log\left(1 + \frac{\left(\sqrt{h_1} + \rho_1(\mathbf{h})\right)^2 P_1(\mathbf{h}) + \left(\sqrt{h_2} + \rho_2(\mathbf{h})\right)^2 P_2(\mathbf{h})}{\sigma_N^2 + \sigma_{N_J}^2(\mathbf{h})}\right)\right] \qquad (106)$$

and $\rho_1(\mathbf{h}), \rho_2(\mathbf{h})$ and $\sigma_{N_J}^2(\mathbf{h})$ are constrained such that

$$E\left[\rho_1^2(\mathbf{h})P_1(\mathbf{h}) + \rho_2^2(\mathbf{h})P_2(\mathbf{h}) + \sigma_{N_J}^2(\mathbf{h})\right] \leq P_J \tag{107}$$

and $\sigma_{N_J}^2(\mathbf{h})$ is nonnegative. The first order KKT conditions for $\rho_1(\mathbf{h})$ and $\rho_2(\mathbf{h})$ are

$$\frac{\sqrt{h_1} + \rho_1(\mathbf{h})}{\sigma_N^2 + \sigma_{N_J}^2(\mathbf{h}) + \left(\sqrt{h_1} + \rho_1(\mathbf{h})\right)^2 P_1(\mathbf{h}) + \left(\sqrt{h_2} + \rho_2(\mathbf{h})\right)^2 P_2(\mathbf{h})} + \lambda \rho_1(\mathbf{h}) = 0 \tag{108}$$

$$\frac{\sqrt{h_2} + \rho_2(\mathbf{h})}{\sigma_N^2 + \sigma_{N_J}^2(\mathbf{h}) + \left(\sqrt{h_1} + \rho_1(\mathbf{h})\right)^2 P_1(\mathbf{h}) + \left(\sqrt{h_2} + \rho_2(\mathbf{h})\right)^2 P_2(\mathbf{h})} + \lambda \rho_2(\mathbf{h}) = 0 \tag{109}$$

which result in

$$\frac{\rho_1(\mathbf{h})}{\sqrt{h_1}} = \frac{\rho_2(\mathbf{h})}{\sqrt{h_2}} \tag{110}$$

Therefore, for the optimal pair of $\rho_1(\mathbf{h})$ and $\rho_2(\mathbf{h})$, we can define $\rho(\mathbf{h})$ and $P(\mathbf{h})$ such that

$$\rho_i(\mathbf{h}) = \sqrt{h_i}\rho(\mathbf{h}), \ i = 1,2 \tag{111}$$

$$P(\mathbf{h}) = h_1 P_1(\mathbf{h}) + h_2 P_2(\mathbf{h}) \tag{112}$$

and write the KKTs for the jammer's best response as

$$\frac{(1 + \rho(\mathbf{h}))}{\sigma_N^2 + \sigma_{N_J}^2(\mathbf{h}) + (1 + \rho(\mathbf{h}))^2 P(\mathbf{h})} + \lambda \rho(\mathbf{h}) = 0 \tag{113}$$

$$-\frac{(1 + \rho(\mathbf{h}))^2 P(\mathbf{h})}{(\sigma_N^2 + \sigma_{N_J}^2(\mathbf{h}))(\sigma_N^2 + \sigma_{N_J}^2(\mathbf{h}) + (1 + \rho(\mathbf{h}))^2 P(\mathbf{h}))} + \lambda + \xi(\mathbf{h}) = 0 \tag{114}$$

where $\xi(\mathbf{h})$ is the complementary slackness variable for $\sigma_{N_J}^2(\mathbf{h})$. From (113) and (114), the best response jamming strategy can be described in terms of $\mathbf{h}$ and $P(\mathbf{h})$. Now, for any pair of user power allocations $P_1(\mathbf{h})$ and $P_2(\mathbf{h})$ and the corresponding jamming best response, the capacity can be written as

$$C = \frac{1}{2}E\left[\log\left(1 + \frac{(1 + \rho(\mathbf{h}))^2 P(\mathbf{h})}{\sigma_N^2 + \sigma_{N_J}^2(\mathbf{h})}\right)\right] \tag{115}$$

Assume that both users transmit at $\mathbf{h} = (h_1, h_2)$. Since the jamming best response is only a function of $\mathbf{h}$ and $P(\mathbf{h})$, the KKTs for the user power allocations can be written as

$$\frac{\mathrm{d}C}{\mathrm{d}P(\mathbf{h})}\frac{\mathrm{d}P(\mathbf{h})}{\mathrm{d}P_1(\mathbf{h})} + \lambda_1 = 0 \tag{116}$$

$$\frac{\mathrm{d}C}{\mathrm{d}P(\mathbf{h})}\frac{\mathrm{d}P(\mathbf{h})}{\mathrm{d}P_2(\mathbf{h})} + \lambda_2 = 0 \tag{117}$$

which, using (112), result in

$$\frac{h_1}{h_2} = \frac{\lambda_1}{\lambda_2} \tag{118}$$

Therefore, $P_1(\mathbf{h})$ and $P_2(\mathbf{h})$ cannot be non-zero at the same time, which means that, at any $\mathbf{h} = (h_1, h_2)$, only one user transmits. Given that only one user is active at any given time, the rest of the two user results immediately follow the results of the single-user case.

## 5 Summary and Conclusion

We characterized the Nash equilibrium solution corresponding to the mutual information game in a non-fading multiple access channel with a correlated jammer. We showed that whether the jammer knows the user signals, or it only has access to a noise corrupted version of the superposition of the user signals, the game has a solution, and the optimal strategies are Gaussian signalling for the users and linear jamming for the jammer.

In fading channels, except for the case when the jammer is correlated and both the user and the jammer have access to the user channel state, we showed that the mutual information game admits a Nash equilibrium solution, and characterized the corresponding user and jammer signalling and power allocation strategies. When the jammer is correlated and both the users and the jammer have access to the channel state, we showed that a set of simultaneously optimal power allocation functions for the users and the jammer does not always exist, and consequently characterized the max-min user power allocation strategies and the corresponding jammer power allocation strategy.

## References

[1] S. Shafiee and S. Ulukus, "Correlated Jamming in Multiple Access Channels," *Conference on Information Sciences and Systems, Baltimore, MD*, March 2005.

[2] S. Shafiee and S. Ulukus, "Capacity of Multiple Access Channels with Correlated Jamming," *Military Communication Conference, Atlantic City, NJ*, October 2005.

[3] M. Médard, "Capacity of Correlated Jamming Channels," *Proc. 35th Annual Allerton Conference on Communications, Control and Computing, Monticello, IL*, September-October 1997.

[4] A. Kashyap, T. Basar and R. Srikant, "Correlated Jamming on MIMO Gaussian Fading Channels," *IEEE Transactions on Information Theory*, 50(9):2119–2123, September 2004.

[5] A. Bayesteh, M. Ansari and A. K. Khandani, "Effect of Jamming on the Capacity of MIMO Channels," *Proc. 42nd Annual Allerton Conference on Communications, Control and Computing, Monticello, IL*, September 2004.

[6] S. Yang, "The Capacity of Communication Channels with Memory," *PhD Thesis*, Harvard University, May 2004.

[7] T. M. Cover, J. A. Thomas, *Elements of Information Theory*, John Wiley & Sons, 1991.

[8] D. Fudenberg, J. Tirole, *Game Theory*, MIT Press, 1991.

[9] E. Burger, *Introduction to the Theory of Games*, Prentice-Hall, 1963.

[10] D. P. Bertsekas, *Convex Analysis and Optimization*, Athena Scientific, 2003.

[11] H. V. Poor, *An Introduction to Signal Detection and Estimation*, 2nd Ed., Springer-Verlag, 1994.

[12] T. W. Anderson, "The Integral of a Symmetric Unimodal Function Over a Symmetric Convex Set and Some Probability Inequalities," *Proceedings of American Mathematical Society*, 6(2):170–176, April 1955.

[13] A. Goldsmith and P. Varaiya, "Capacity of Fading Channels with Channel Side Information," *IEEE Transactions on Information Theory*, 43(6):1986–1992, November 1997.

[14] G. Caire and S. Shamai, "On the Capacity of Some Channels with Channel State Information," *IEEE Transactions on Information Theory*, 45(6):2007–2019, September 1999.

[15] R. Knopp and P. A. Humblet, "Information Capacity and Power Control in Single-cell Multiuser Communications," *IEEE International Conference on Communications, Seattle, WA*, June 1995.

[16] R. J. McEliece, "Communication in the Presence of Jamming-An Information Theoretic Approach," in *Secure Digital Communications, CISM Courses and Lectures*, no. 279, G. Longo, Ed. New York: Springer Verlag, 1983.

[17] E. Biglieri, J. Proakis and S. Shamai, "Fading Channels: Information-Theoretic and Communication Aspects," *IEEE Transactions on Information Theory*, 44(6):2619–2692, October 1998.

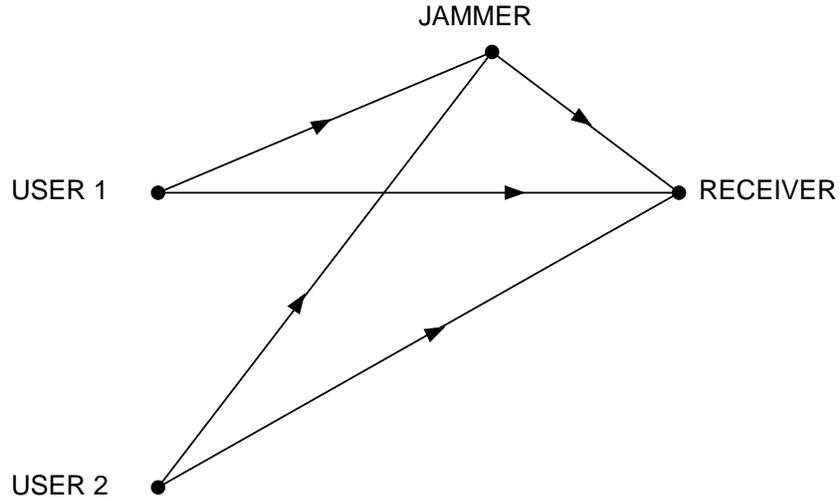

Figure 1: A communication system with two users and one jammer.

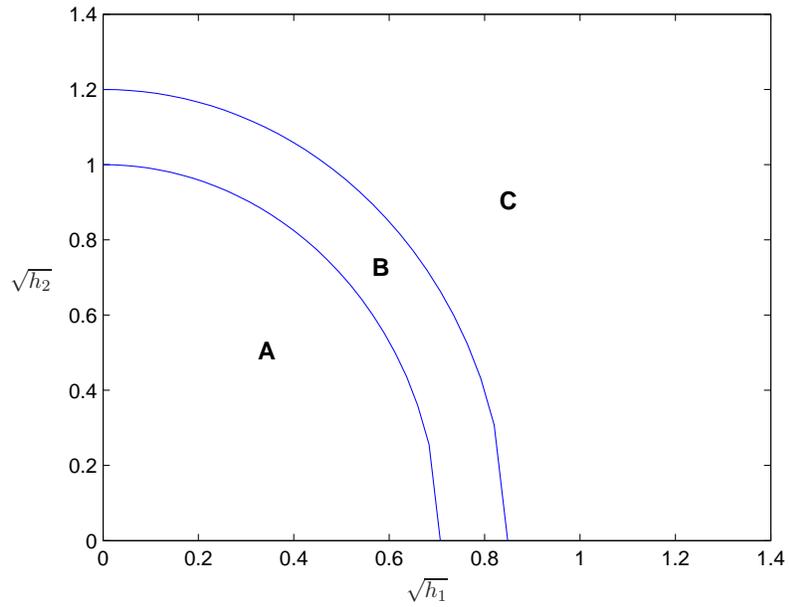

Figure 2: Jamming decision regions when $\gamma = 1$, $P_1 = 10$, $P_2 = 5$, $P_J = 5$ and $\sigma_N^2 = 1$.

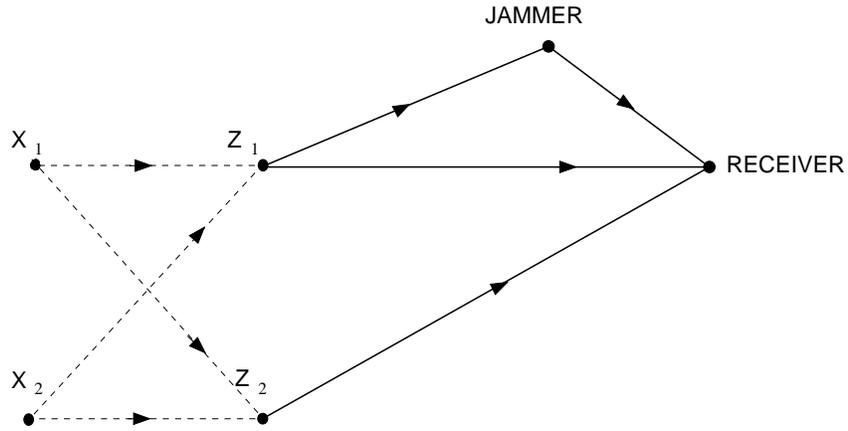

Figure 3: An interpretation of a communication system with two users and one jammer with eavesdropping information.

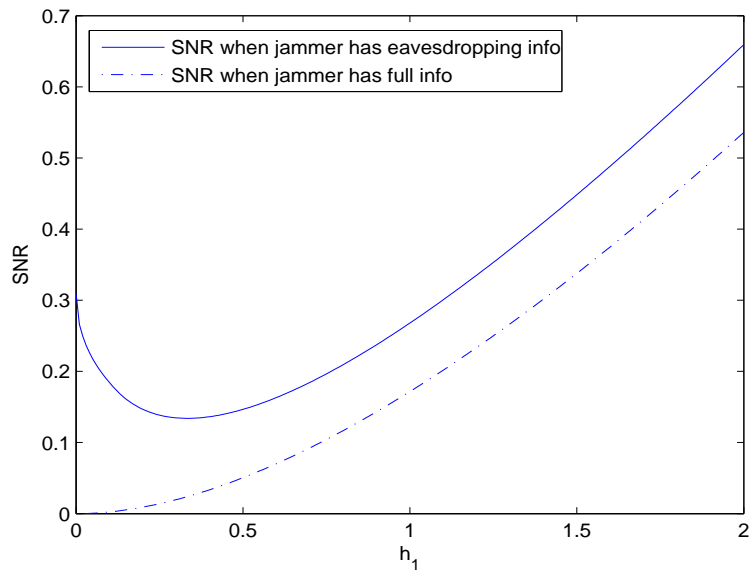

Figure 4: SNR as a function of $h_1$ when $h_2 = g_1 = g_2 = \gamma = 1$ and the powers are $P_1 = P_2 = P_J = \sigma_N^2 = 1$, for the cases when the jammer has full information and when it has eavesdropping information.

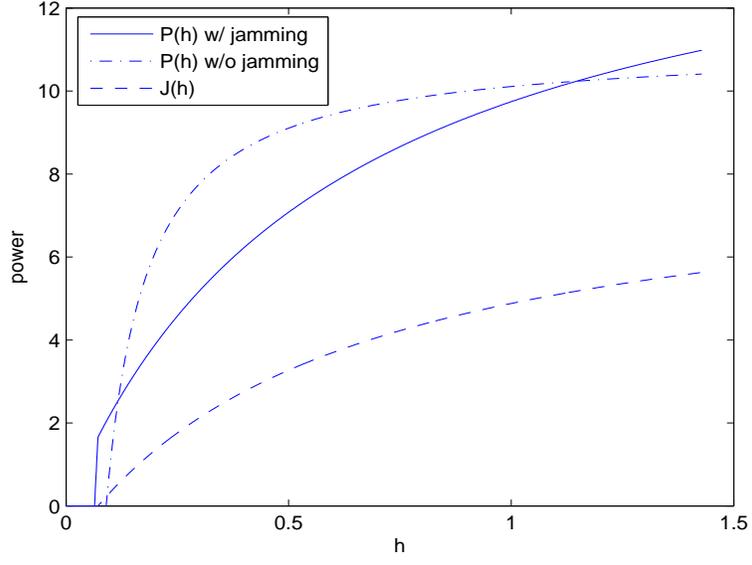

Figure 5: $P(h)$ and $J(h)$ for $E[P(h)] = 10$, $E[J(h)] = 5$ and $\sigma_N^2 = 1$.

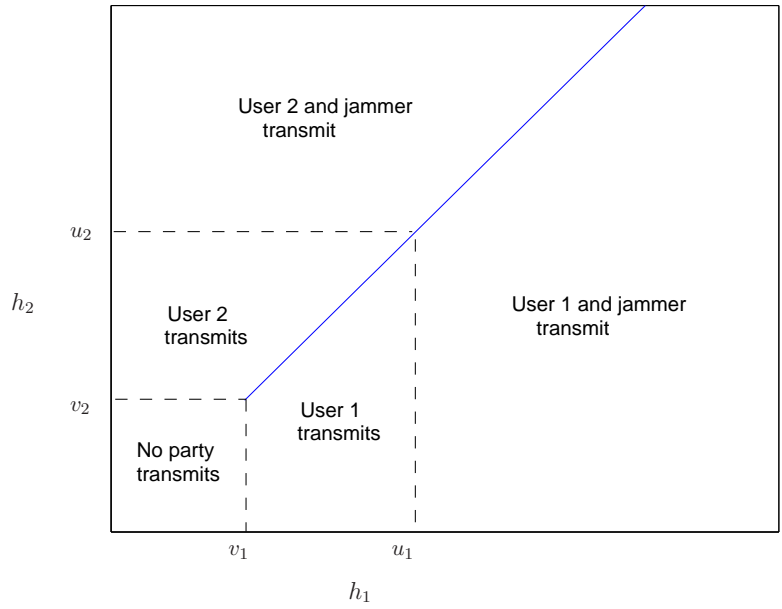

Figure 6: User/jammer transmission regions in uncorrelated jamming with CSI.

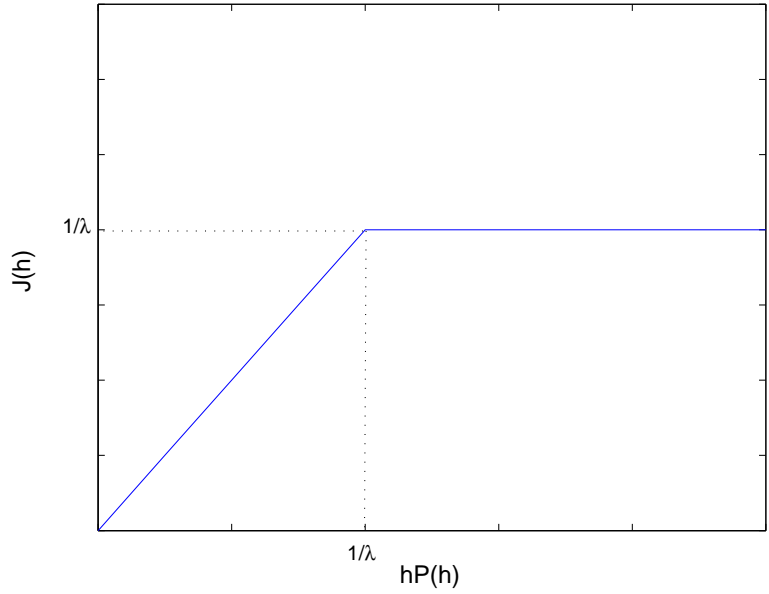

Figure 7: Jammer best response power allocation in correlated jamming with CSI.

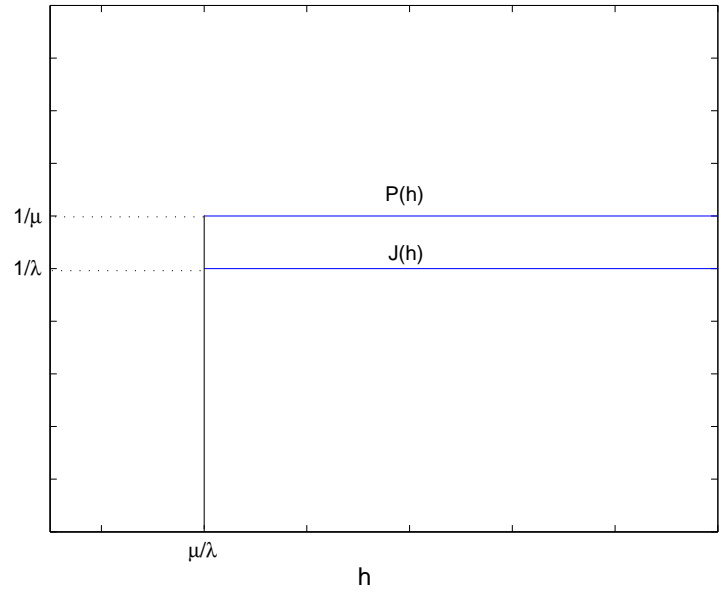

Figure 8: Max-min user power allocation and the corresponding jammer best response power allocation in correlated jamming with CSI.

30